\documentclass[prb,aps,amsmath,showpacs,preprint]{revtex4}

\usepackage{graphicx}% Include figure files

\usepackage{dcolumn}% Align table columns on decimal point

\usepackage{bm}% bold math

\begin{document}
\title{Valency configuration of transition metal impurities in ZnO} 
\author{L. Petit$^{1}$, T. C. Schulthess$^{1}$, A. Svane$^{2}$, W.M. Temmerman$^{3}$, Z. Szotek$^{3}$
, and A. Janotti$^{4}$}
\affiliation{$^{1}$Computer Science and Mathematics Division, and Center for Computational Sciences, 
Oak Ridge National Laboratory, Oak Ridge, TN 37831, USA\\
$^{2}$ Department of Physics and Astronomy, University of Aarhus, 
DK-8000 Aarhus C, Denmark\\
$^{3}$ Daresbury Laboratory, Daresbury, Warrington WA4 4AD, UK\\
$^{4}$ Materials Department, University of California, Santa Barbara, CA 93106-5050}
\date{\today}

\begin{abstract}
We use the self-interaction corrected local spin-density approximation to investigate the ground state valency 
configuration of transition metal (TM=Mn, Co) impurities in n- and p-type ZnO. 
We find that in pure Zn$_{1-x}$TM$_x$O, the localized TM$^{2+}$ configuration is energetically favoured over 
the itinerant $d$-electron configuration
of the LSD picture. Our calculations indicate furthermore, that
the (+/0) donor level is situated in the ZnO gap. Consequently,
 for n-type conditions, with the Fermi energy $\epsilon_F$ close to the conduction band minimum,
TM remains in the 2+ charge state, whilst for p-type conditions, with $\epsilon_F$ close to the valence band maximum,
the 3+ charge state is energetically preferred. 
In the latter scenario, modelled here by co-doping with N, the additional delocalized $d$-electron charge 
transfers into the hole states at the top 
of the valence band, and hole carriers will only exist, if the N concentration exceeds the TM impurity
concentration.
 
\end{abstract}
\pacs{}
\maketitle

%\narrowtext

Based on the possible interplay between electronic properties and spin-functionality, 
diluted magnetic semiconductors (DMS) are expected to play a major part in the development of 
the next generation of electronic devices.~\cite{wolf} The considerable challenge consists in 
designing  materials that remain ferromagnetic above room temperature. In Mn doped GaAs, where 
ferromagnetism is well established, the currently highest achieved Curie temperature 
is T$_C$ $\sim$ 170 K. Ferromagnetism at even higher temperatures has been reported in some semiconductors 
(for a review, see for example Pearton {\it et al.}~\cite{pearton}), but doubts persist as to the 
carrier induced nature of the observed magnetic order.
Recently, the prediction by Dietl {\it et al.}~\cite{dietl0} of room temperature ferromagnetism 
in p-type Zn$_{1-x}$Mn$_x$O has generated considerable research activity, both in theory and experiment.  
So far, no conclusive experimental evidence has emerged that could either confirm or disprove the prediction. 
Various experimental investigations of the magnetic order in Zn$_{1-x}$TM$_x$O give contradictory results, 
ranging from spin glass~\cite{fukumura} and antiferromagnetic behaviour~\cite{yoon,risbud} to ferromagnetism 
below~\cite{jung} or even above~\cite{sharma,sluiter,lin} room temperature. It has also been suggested that the 
observed ferromagnetism might be due to the formation of secondary manganese oxide phases.~\cite{kim} 
Generally, the observed magnetic properties of the DMS seem to depend strongly on the method by which they are synthesized.
From all these conflicting results, the overall picture (though not universally accepted) seems to emerge, 
that Zn$_{1-x}$TM$_x$O
will not be ferromagnetic,~\cite{lawes} unless additional hole or electron carriers are co-doped. 
Comprehensive overviews on the current state of research into ZnO spintronics can be found in references
[\onlinecite{fukrev1,fukrev2,peartrev,janisch}].   

From the theory point of view, there are currently two main approaches to modelling dilute magnetic semiconductors
that differ rather fundamentally in the way the 
transition metal $d$-electrons are described. In the Zener model approach proposed by Dietl {\it et al.}~\cite{dietl0}
it is assumed that
the on-site $d$-$d$ correlations are significant enough to prevent hopping, causing the transition metal $d$-electrons to remain
localized on each their sites. 
%In this scenario, it is argued that any observed ferromagnetism originates from the 
%RKKY-like interaction between the localized transition metal moments, and additional delocalized hole carriers in the compound.
In the band model approach it is assumed that the gain in kinetic energy associated with $d$-electron delocalization 
is large enough to overcome the on-site correlations,
and that consequently the $d$-electrons fill up the resulting exchange split $d$-band, situated in the case of ZnO in the
middle of the gap. 
%Ferromagnetism if it exists is in this scenario based on the double-exchange interaction.
Both the Zener model description,~\cite{dietl0} and the 
band-electron picture~\cite{sato,spaldin} have been 
put forward to predict high temperature (even room temperature) ferromagnetism in TM doped p-type ZnO. 

Despite this apparent agreement on ferromagnetism, the two models nevertheless differ with respect to the proposed mechanism, 
which ultimately
can be traced back to the very different assumptions regarding the ground state valency configuration of the TM impurities. 
%In order to determine the actual $d$-electron ground state configuration, one needs to be able to compare the total energies of the two scenarios.
With the self-interaction corrected (SIC)-LSD approximation, used in the present work,~\cite{svane,temmerman} 
delocalized and localized $d$-electrons are treated on an equal footing, by removing an unphysical
self-interaction, inherent in LSD, for every localized d-electron 
.~\cite{perdew} Thus the localized mean field scenario, and the delocalized bandstructure
scenario can be compared, and the ground state configuration of the TM-ion can be deduced from the global energy minimum.

\section{SIC-LSD method}

Whereas density functional theory (DFT) is in principle exact, the very popular local spin density approximation (LSD) 
used to describe exchange and correlation introduces an unphysical self-interaction. 
This can be demonstrated on a single electron moving in an external potential, which should be described
by the total energy functional
\begin{equation}
E^{DFT}[n_{\alpha}]=E_{kin}[n_{\alpha}]+V_{ext}[n_{\alpha}]
\end{equation}
but which in the LSD approximation is given by 
\begin{equation}
E^{LSD}[n_{\alpha}]=E_{kin}[n_{\alpha}]+V_{ext}[n_{\alpha}]+U_{H}[n_{\alpha}]+E_{xc}^{LSD}[n_{\alpha}].
\end{equation}
Here, U$_{H}$[n$_{\alpha}$] and E$_{xc}^{LSD}$[n$_{\alpha}$] are respectively the self-Hartree and self-exchange-correlation
energies, for the single electron density [n$_{\alpha}$], and together defining the self-interaction as 
\begin{equation}
\delta_{\alpha}^{SIC}=U_{H}[n_{\alpha}]+E_{xc}^{LSD}[n_{\alpha}].
\end{equation}
In the solid, using the SIC-LSD method, the self-interaction for each occupied orbital is subtracted from the LSD functional
\begin{equation}
E^{SIC}[\{\psi_{\alpha}\}]=E^{LSD}[n]-\sum_{\alpha}^{occ}\delta_{\alpha}^{SIC},
\end{equation}
where the total energy $E^{SIC}[\{\psi_{\alpha}\}]$ is expressed as functional of the set of orbitals $\psi_{\alpha}$,
and $n$ is the total charge density of the system. For states of Bloch symmetry in an infinite crystal, i.e.
band states, the self-interaction vanishes. To benefit from the self-interaction correction, a localized state must be formed.
The ensuing many-electron wavefunction becomes a mixture of Bloch-type and Heitler-London type one particle states,
i.e. delocalized and localized d-electrons coexist. The delocalized $d$-states  together with the remaining valence
electrons continue to be treated in the LSD approximation, whilst those $d$-states that are treated as localized
gain the self-interaction (localization) energy, but loose the corresponding band formation energy.
TM configurations with varying numbers of localized $d$-states - in general different 'localization scenarios' - can 
be explored, and compared with respect to their total energies. Consequently, the preferred ground state 
configuration, as far as the number of localized $d$-electrons is concerned, can be determined from the global 
energy minimum.

Different configurations of localized and delocalized $d$-electrons will give different nominal valencies, which in the
SIC-LSD are defined as: $N_{val}=Z-N_{core}-N_{SIC}$, where $Z$ is the atomic charge, $N_{core}$ is the number of 
core and semicore electrons, and $N_{SIC}$ is the number of self-interaction corrected $d$-electrons respectively. According
to this definition, a divalent TM valency refers to the localized $d^5$ configuration of the Mn ion (Mn$^{2+}$)
, and the localized $d^7$
Co ion (Co$^{2+}$) respectively. The trivalent configuration has one less $d$-electron localized, and thus  refers to the localized 
$d^4$ Mn ion (Mn$^{3+}$)
and $d^6$ Co ion (Co$^{3+}$). The LSD picture corresponds to the scenarios with all $d$-electrons delocalized,
i.e. the $d^0$ configuration for the Mn ion (Mn$^{7+}$) and the Co ion (Co$^{9+}$).     

\section{Band picture versus localized picture}

\begin{figure}
\includegraphics[scale=0.40,angle=270]{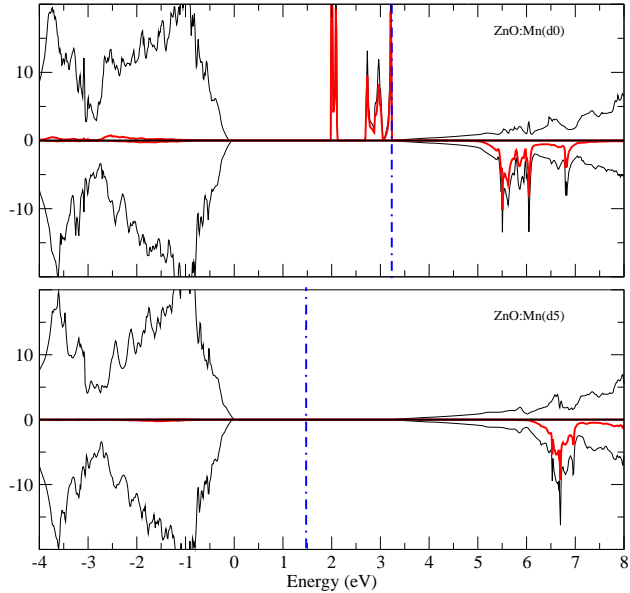}
\caption{Total (black), and Mn-$d$ projected (red) DOS as a function of energy, in states per eV, of Zn$_{1-x}$Mn$_x$O a) in the delocalized (band picture) Mn$^{7+}$ configuration and b) in the 
localized divalent Mn$^{2+}$ configuration. } 
\label{mn5d0d}
\end{figure}

We use the SIC-LSD method to study the validity of the two above mentioned electronic structure models of TM impurities in ZnO, and to determine which of the
two assumes the correct TM ground state configuration, based on total energy considerations. 
When doping ZnO with Mn/Co, the TM ions occupy the Zn sites without changing the wurtzite structure.~\cite{fukumura1} 
In our calculations, Zn$_{1-x}$TM$_x$O is modelled, by substituting a single Zn in a (2x2x2) supercell consisting of 
16 ZnO formula units, with a TM impurity. 
The SIC-LSD scheme is implemented within the all-electron tight-binding linear-muffin-tin orbitals (TB-LMTO) method, with the 
atomic sphere approximation (ASA). Spin-orbit coupling is fully included in the self-consistency cycles. 
Empty spheres are introduced on high symmetry interstitial sites. 
The electron charge density was sampled using 27 $k$-points
in the irreducible wedge of the Brillouin zone. All calculations use the experimental lattice parameter. 
The resulting densities of states (DOS) for the delocalized $d$-electron scenario (LSD picture), and the localized $d^5$ scenario 
are shown in Fig. \ref{mn5d0d}a and Fig. \ref{mn5d0d}b respectively.

In Fig. \ref{mn5d0d}a, the delocalized electrons fully occupy the spin-up channel of the exchange split $d$-band whilst the spin-down band remains
empty. The proponents of the band picture conclude that ferromagnetism, due to double-exchange, can not occur in this scenario. The same
argument however is less useful for explaining the absence of ferromagnetism in Zn$_{1-x}$Co$_x$O, where the two 
additional $d$-electrons start filling up the spin-down band, and the corresponding LSD DOS is reminiscent of a half-metal. 
In Fig. \ref{mn5d0d}b, the 5 Mn $d$-electrons have been SI-corrected, and are therefore no longer available for band formation. Instead they form localized 
levels at energies below the valence band maximum (VBM). Note that the SIC-LSD, like any other density-functional theory, 
does not provide excited state properties. In particular, the approach cannot accurately determine the removal energies
of localized states, and the bare SI-corrected d-bands always appear at too high binding energies due to the neglect of screening
and relaxation effects,~\cite{where are the d-states} and are therefore not shown in the DOS plots. 
The Fermi level is situated in the middle of the ZnO gap, and there are
no free carriers, which for the proponents of the localized model explains the absence of carrier mediated ferromagnetism.
The corresponding DOS for Zn$_{1-x}$Co$_x$O is quite similar to Fig. \ref{mn5d0d}b, but with seven Co $d$-states now SI-corrected in the localized 
scenario.
Comparing the total energies for both models (columns 2 and 5 in Table I), we find that the localized Mn$^{2+}$ and Co$^{2+}$ scenarios are energetically more favourable 
than the corresponding band scenarios, by respectively 3 eV for Zn$_{1-x}$Mn$_x$O (row 2) and 5.5 eV for Zn$_{1-x}$Co$_x$O (row 8). 
\begin{table}[b]
\label{localize}
\caption{Total energies in eV of Mn/Co doped p- and n-type ZnO.  }
\begin{ruledtabular}
\begin{tabular}{|r|r|r|r|r|}
Compound & Mn$^{2+}(d^5) $ & Mn$^{3+}(d^4)$ & Mn$^{4+}(d^3)$  & Mn$^{7+}(d^0)$ \\
\hline
Zn$_{15/16}$Mn$_{1/16}$O &{\bf -113.13}&-112.61&-112.03&-110.12 \\
\hline
Zn$_{15/16}$Mn$_{1/16}$O$_{15/16}$N$_{1/16}$ &-111.40&{\bf -113.70}&-112.58& \\
\hline
Zn$_{15/16}$Mn$_{1/16}$O$_{14/16}$N$_{2/16}$ &-110.80&{\bf -111.48}&-111.35& \\
\hline
Zn$_{14/16}$Mn$_{1/16}$Ga$_{1/16}$O &{\bf -115.91}& -115.40&& \\
\hline
Zn$_{14/16}$Mn$_{1/16}$Sn$_{1/16}$O &{\bf -110.49}& -109.89&& \\
\hline
\hline
 & Co$^{2+}(d^7) $ & Co$^{3+}(d^6)$ & Co$^{4+}(d^5)$  & Co$^{9+}(d^0)$ \\
\hline
Zn$_{15/16}$Co$_{1/16}$O &{\bf -114.04}&-113.40&-112.70&-108.38 \\
\hline
Zn$_{15/16}$Co$_{1/16}$O$_{15/16}$N$_{1/16}$ &-112.31&{\bf -113.90}&-113.25& \\
\hline
Zn$_{15/16}$Co$_{1/16}$O$_{14/16}$N$_{2/16}$ &-110.28&{\bf -112.01}&-111.79& \\
\hline
Zn$_{14/16}$Co$_{1/16}$Sn$_{1/16}$O &{\bf -111.48}& -110.89&& \\
\hline
\end{tabular}
\end{ruledtabular}
%\caption{Total energies in eV of Mn/Co doped p- and n-type ZnO.  }
\end{table}

\begin{figure}
\includegraphics[scale=0.40,angle=270]{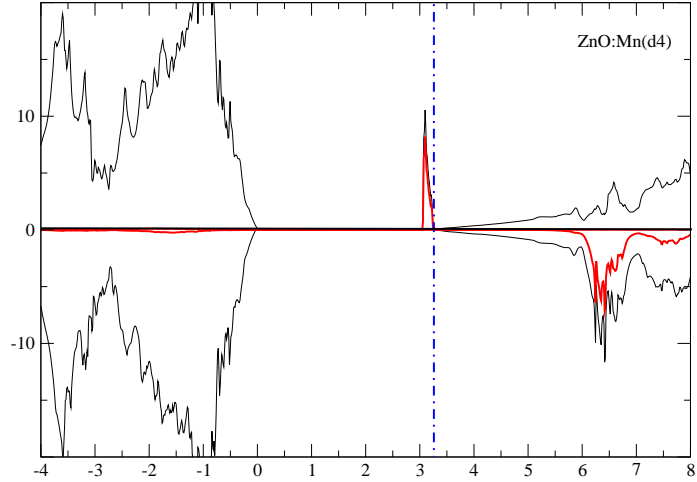}
\caption{Total (black), and Mn-$d$ projected (red) DOS, in states per eV, as a function of energy, in eV, 
of Zn$_{1-x}$Mn$_x$O in the trivalent Mn$^{3+}$ configuration. } 
\label{mn4d}
\end{figure}

The band picture is clearly not preferred energetically, relative to the localized picture, however we still need to investigate whether
the global energy minimum is obtained for an intermediate valency configuration of coexisting localized and delocalized $d$-electrons. 
In Fig. \ref{mn4d} the DOS of Zn$_{1-x}$Mn$_x$O is shown with the Mn ion in the Mn$^{3+}$ configuration, i.e.
only four of the $d$-electrons are localized by the SIC. The remaining $d$-electron, which is now treated in the LSD approximation,
fills up the small $d$-peak situated in the middle
of the ZnO gap. From Table I (row 2, columns 2 and 3), we see that this configuration is energetically less favourable than the TM$^{2+}$ configuration. 
Similarly for Co doped ZnO the global energy minimum is obtained in the Co$^{2+}$ scenario (Table I, row 8), and we conclude
that both Zn$_{1-x}$Mn$_x$O and Zn$_{1-x}$Co$_x$O prefer the localized $d$-electron ground state configuration. 
The TM$^{2+}$ configuration of transition metal impurities in ZnO is in agreement with experiment,~\cite{dorain,wi} and in support of
the electronic structure underlying the Zener model, rather than the standard band picture.

\begin{figure}
\includegraphics[scale=0.40,angle=0]{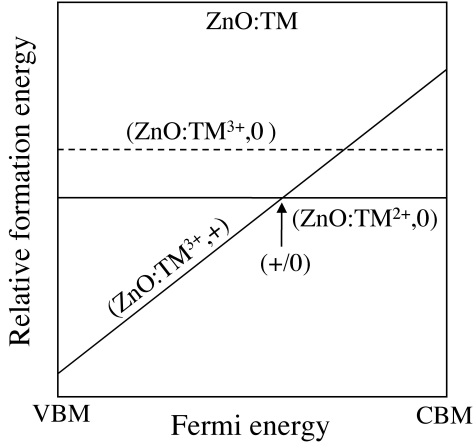}
\caption{Schematic plot of the formation energy as a function of Fermi energy for ZnO:TM, illustrating the link between the localization
and delocalization picture and the TM donor ionization level in the band-gap. Only the (+/0) level between the positively charged
TM$^{3+}$ and the neutral TM$^{2+}$ valence configurations is indicated.}
\label{doping}
\end{figure}

In Fig. \ref{mn4d}, the gap state that forms when treating an additional $d$-electron as band electron, is relatively narrow, and completely filled. The
corresponding gain in band formation energy is insufficient to overcome the loss in SI-energy, and the $d$-electron prefers to remain localized. If
however the Fermi level could be positioned below this gap state, for example through additional acceptor dopants, one could very well imagine that
the resulting gain in energy from charge transfer to the Fermi level would be large enough for the delocalization to occur. The dependence on Fermi level
position can be expressed through the formation energy 
\begin{equation}
E_f(ZnO:TM^v,q)=E_{tot}(ZnO:TM^v,q)-E_{tot}(ZnO)-\mu_{Zn}+\mu_{TM}+ q \epsilon_F
\end{equation}
where (ZnO:TM$^v$,q) refers to an impurity of valency $v$, and charge $q$, $\mu_{Zn}$ and $\mu_{TM}$ are the respective chemical potentials, $\epsilon_F$
is the Fermi energy, and E$_{tot}$ the total energy of the system with impurity. 
Three scenarios need to be considered: In the neutral charge states, the substitutional impurity can either assume the divalent configuration,
TM$^{2+}$ $\equiv$ (ZnO:TM$^{2+}$,0), which does not introduce states in the gap (Fig. \ref{mn5d0d}), or the trivalent configuration 
TM$^{3+}$ $\equiv$ (ZnO:TM$^{3+}$,0) which has an electron $d$-state in the gap (Fig. \ref{mn4d}).  In the charged scenario (ZnO:TM$^{3+}$,+)
the delocalized $d$-electron has been transferred to the conduction or acceptor band, ionizing the TM impurity into the positive charge state.
The formation energy as a function of Fermi level
is plotted in Fig. \ref{doping}. The two horizontal lines correspond to the neutral scenarios, that are independent of the Fermi level
position. The formation energy
for the charged scenario changes linearly with Fermi level, as follows from equation (5), and as indicated by the skew line. 
The delocalization process
\begin{equation}
TM^{2+} \rightarrow TM^{3+} + electron_{\epsilon_F}
\end{equation}
will go ahead, if the position of the Fermi energy is such that E$_f$(ZnO:TM$^{3+}$,+) $\leq$ E$_f$(ZnO:TM$^{2+}$,0).
The donor transition
level (+/0) is defined by the Fermi energy above which the TM impurity is in the divalent state, and below which it is in the positively
charged trivalent state. 
From SIC-LSD calculations, we have determined that the donor level in Zn$_{1-x}$TM$_x$O is positioned more or less centrally in the 
ZnO gap~\cite{petitzno}. 
The same conclusion can be drawn qualitatively from applying the numbers of Table I to Fig. \ref{doping}: Since the neutral 
trivalent state (ZnO:TM$^{3+}$,0) has
a gap state in the high energy part of the band gap, and since it is situated only $\sim$ 0.5 eV above the neutral 
divalent (ZnO:TM$^{2+}$,0) configuration,
the donor level (+/0) will be situated in the 3.4 eV wide ZnO gap, as is also found experimentally.~\cite{madelung}

\begin{figure}
\includegraphics[scale=0.40,angle=270]{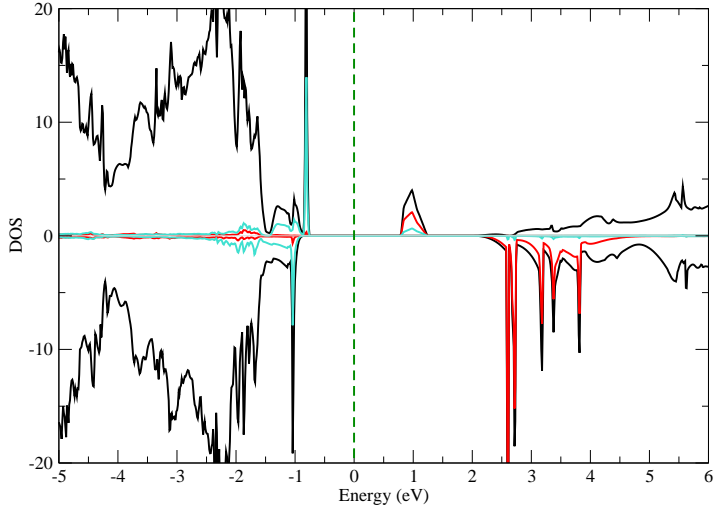}
\caption{Total (black),  N-$p$(blue), and Mn-$d$(red) DOS as a function of energy, in states per eV, of Zn$_{15/16}$Mn$_{1/16}$O$_{15/16}$N$_{1/16}$ in the trivalent Mn$^{3+}$ 
ground state configuration.}
\label{znomnn}\end{figure}

To investigate the actual effect of additional dopants on the electronic structure of Zn$_{1-x}$TM$_x$O, we use the SIC-LSD to determine the ground state configuration
for both  p-  and n-type scenarios. Acceptor (p-type) doping is achieved by replacing a single O in the supercell by N, whereas n-type doping is modelled
by replacing a single Zn atom by either Ga or Sn. The total energy results for the different dopants and different TM valency configurations are
given in Table I. We find that, with N as additional dopant, the trivalent TM configuration becomes energetically most favourable, 
both for Mn (row 3) and Co 
(row 9). Nitrogen, having one p-electron less than O, acts as acceptor when introduced into ZnO, with the corresponding acceptor level situated near the top
of the valence band (Fig. \ref{znomnn}). Given that the (+/0) level is situated above the acceptor level, the delocalization process (6) can go ahead, 
leaving behind an empty $d$-band (gap state). The energy gain from charge transfer and hybridization in the TM$^{3+}$ configuration
is obviously large enough to overcome the corresponding loss in SIC (localization) energy.  

In their paper on the possibility of magnetism in Mn doped ZnO, Dietl {\it et al.}~\cite{dietl0} state that with n-type doping one 
can not expect ferromagnetic order, 
except maybe for very low temperature, but they predict room temperature ferromagnetism in the p-type scenario, where the hole carriers introduced
by co-doping with N will mediate the interaction. The assumption is that the ground state configuration remains Mn$^{2+}$,
 based on the argumentation that the
(+/0) level is situated below the VBM.~\cite{dietl1} This picture is thus in contrast to our calculation, which finds the (+/0) level situated above the N based acceptor 
states, and results in the Mn$^{3+}$ ground state configuration. The importance of determining the correct ground state configuration
becomes clear from Fig. \ref{znomnn}, where the actual trivalent configuration implies charge transfer into the acceptor states, 
and absence of hole carriers, while 
a hypothetical divalent scenario with five localized $d$-states would imply absence of the gap state and charge transfer, leaving the N acceptor states
partially unfilled (Dietl's scenario). 
From the SIC-LSD calculations it follows that, the band description of the $d$-states does not fully account for 
the correct electronic structure of the Zn$_{1-x}$TM$_x$O ground state, even when co-doped with N. But also the Zener model, 
with hole carriers mediating the magnetic interaction between localized spins residing on TM$^{2+}$ ions, 
is not a true representation of the ground state. There are no hole carriers in either Zn$_{15/16}$Mn$_{1/16}$O$_{15/16}$N$_{1/16}$ 
or Zn$_{15/16}$Co$_{1/16}$O$_{15/16}$N$_{1/16}$ and carrier mediated ferromagnetism can therefore not occur.

It seems plausible that increasing the concentrations of the dopants, [TM] and [N], will not change the overall picture of 
fully compensated acceptor states, as long as [TM]=[N]. However, the qualitative picture changes considerably if we increase 
the relative amount of N impurities, i.e., if [N]$>$[TM]. Substituting two of the O atoms by N, but with a single TM impurity in the 
32 atom ZnO supercell, we find that, contrary to what one might expect, overdoping with N does not result in a further delocalization, 
i.e., a transition from TM$^{3+}$ $\longrightarrow$ TM$^{4+}$ + e$^-$. From Table I we see that the TM$^{3+}$ configuration gives the 
lowest total energy, both for Zn$_{15/16}$Mn$_{1/16}$O$_{14/16}$N$_{2/16}$ (row 4), and 
Zn$_{15/16}$Co$_{1/16}$O$_{14/16}$N$_{2/16}$ (row 10). The additional N impurity is thus not compensated by TM $d$-electrons 
which results in an impurity band at the top of the valence band, which is now only partially filled.  

\begin{figure}
\includegraphics[scale=0.40,angle=0]{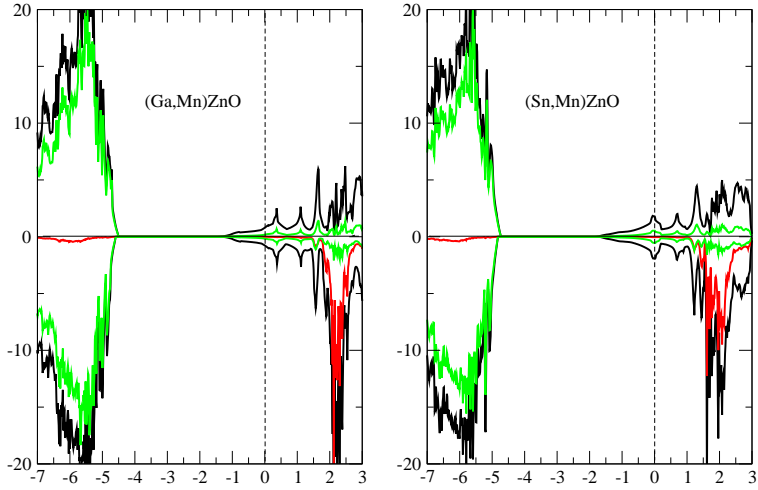}
\caption{Total DOS, in states per eV, as a function of energy, in eV, of a) Zn$_{14/16}$Mn$_{1/16}$Ga$_{1/16}$O and b) Zn$_{14/16}$Mn$_{1/16}$Sn$_{1/16}$O, in the divalent Mn$^{2+}$ configuration. The green, red, and black lines respectively represent the O-$p$ projected, Mn-$d$ projected, and total DOS's. The Fermi level (zero energy)
is now situated in the conduction bands.}
\label{ntype}\end{figure}

Co-doping with N has revealed itself to be rather difficult,~\cite{joseph} which according to calculations by 
Yamamoto {\it et al.}~\cite{yamamoto} can be related to the corresponding increase in the Madelung energy. 
Co-doping with electron donors, on the other hand, can be readily achieved. The theoretical prediction, that high 
temperature ferromagnetism will not occur in n-type Mn-doped ZnO, seems to be confirmed by experiment with respectively 
Al and Ga as co-dopants, however ferromagnetism has been reported in Zn$_{1-x}$Mn$_x$O, when co-doped with Sn.~\cite{norton} 
In the case of n-type doping,
the substitution of Zn by either Ga or Sn introduces respectively one or two additional electrons into ZnO, which results in 
an impurity band situated below the conduction band minimum (Fig. \ref{ntype}). From Table I, rows 5, 6, and 11,
 we see that in this case the 
divalent TM configuration remains energetically more favourable, i.e. given that the Fermi level is here 
situated above the (+/0) level, delocalization does not occur, and respectively five (Mn) or seven (Co) $d$-electrons 
remain localized at each of the respective sites. With or
without n-type co-dopant the divalent state (of both Mn and Co)
invariably is 0.5-0.6 eV more favorable than the trivalent one, i.e.
the n-type doping does not influence the TM configuration at all.

With respect to the calculated magnetic moments of the TM impurities in the various scenarios and valency configurations, we find
that the orbital moment remains largely quenched. With no additional co-dopants, i.e. rows 2 and 7 of table I, the spin moments
are approximately 4.5 $\mu_B$ and 2.7 $\mu_B$ for Mn and Co impurities respectively. These moments remain virtually unchanged 
regardless of the considered
configuration, indicating for example that the SIC localized $d$-state of the Mn$^{2+}$ configuration has the same moment as the filled
narrow $d$-band
state of the corresponding Mn$^{3+}$ configuration. 
This situation changes when N is codoped (rows 3 and 9 of Table I). Here in in the TM$^{2+}$ scenario, the moments are 
respectively 4.35 $\mu_B$ for Mn in Zn$_{15/16}$Mn$_{1/16}$O$_{1/16}$N$_{1/16}$ and 2.53 $\mu_B$ for Co in
Zn$_{15/16}$Co$_{1/16}$O$_{1/16}$N$_{1/16}$. However for the corresponding TM$^{3+}$ configuration, 
a $d$-electron has charge transfered into the hole state, with the moment being determined by the remaining SIC localized d-states,
i.e. Mn($d^4$) and Co($d^6$) .
As a result, the magnetic moment of Mn decreases to 3.7 $\mu_B$, and the magnetic moment of Co increases to 3.0 $\mu_B$.

\section{Conclusion}
In summary, we have studied the electronic structure and different valency configurations
of Co and Mn impurities in p- and n-type ZnO using the SIC-LSD {\it ab initio} method. 
From total energy considerations we find that the TM $d$-states remain localized if no additional hole donors are present. The TM$^{2+}$
remains the ground state configuration under n-type conditions. The TM$^{3+}$ becomes more favourable in $p$-type ZnO,
and the holes required for carrier mediated ferromagnetism occur  
only when the N co-dopant concentration exceeds the concentration of the TM impurities.

\begin{acknowledgments}
This work was supported in part by the Defense Advanced Research Project Agency and by the Division of Materials Science and 
Engineering, US Department of Energy. Oak Ridge National Laboratory is managed by UT-Batelle, LLC, for the US Department of Energy 
under Contract No. DE-AC05-00OR22725. The calculations were carried out at the Center for Computational Sciences at Oak Ridge National 
Laboratory, and at the Danish Center for Scientific Computing.
\end{acknowledgments}

%\end{bibliography}

\newpage

%\begin{figure}
%\includegraphics[scale=0.40,angle=0]{znmno.eps}
%\caption{Total DOS of Zn$_{1-x}$Mn$_x$O in the divalent Mn$^{2+}$ configuration (main graph), and the delocalized
%Mn$^{7+}$ configuration (inset).} 
%\label{znmno}
%\end{figure}

\end{document}